\pdfoutput=1
\documentclass[sigconf]{acmart}

\AtBeginDocument{%
  \providecommand\BibTeX{{%
    \normalfont B\kern-0.5em{\scshape i\kern-0.25em b}\kern-0.8em\TeX}}}

\setcopyright{acmcopyright}
\copyrightyear{2024}
\acmYear{2024}
\acmDOI{XXXXXXX.XXXXXXX}

\acmConference[Conference acronym 'XX]{Make sure to enter the correct
  conference title from your rights confirmation emai}{June 03--05,
  2024}{Woodstock, NY}
\acmPrice{15.00}
\acmISBN{978-1-4503-XXXX-X/18/06}

\usepackage{makecell}
\usepackage{enumitem}
\usepackage{multirow}

\newcommand{\ie}{\emph{i.e., }}
\newcommand{\eg}{\emph{e.g., }}

\begin{document}

\title{A Survey of Generative Search and Recommendation in the Era of Large Language Models}

\author{Yongqi Li}
\authornote{Both authors contributed equally to this research.}
\email{liyongqi0@gmail.com}
\affiliation{%
  \institution{The Hong Kong Polytechnic University}
    \streetaddress{}
  \city{}
  \country{}
}

\author{Xinyu Lin}
\authornotemark[1]
\email{xylin1028@gmail.com}
\affiliation{%
  \institution{National University of Singapore}
    \streetaddress{}
  \city{}
  \country{}
}

\author{Wenjie Wang}
\email{wenjiewang96@gmail.com}
\affiliation{%
  \institution{National University of Singapore}
    \streetaddress{}
  \city{}
  \country{}
}

\author{Fuli Feng}
\email{fulifeng93@gmail.com}
\affiliation{%
  \institution{University of Science and Technology of China}
    \streetaddress{}
  \city{}
  \country{}
}

\author{Liang Pang}
\email{pangliang@ict.ac.cn}
\affiliation{%
 \institution{Chinese Academy of Sciences}
    \streetaddress{}
  \city{}
  \country{}
 }

\author{Wenjie Li}
\email{cswjli@comp.polyu.edu.hk}
\affiliation{%
 \institution{The Hong Kong Polytechnic University}
    \streetaddress{}
  \city{}
  \country{}
 }

\author{Liqiang Nie}
\email{nieliqiang@gmail.com}
\affiliation{%
 \institution{Harbin Institute of Technology (Shenzhen)}
    \streetaddress{}
  \city{}
  \country{}
 }

\author{Xiangnan He}
\email{xiangnanhe@gmail.com}
\affiliation{%
 \institution{University of Science and Technology of China}
    \streetaddress{}
  \city{}
  \country{}
 }

\author{Tat-Seng Chua}
\email{dcscts@nus.edu.sg}
\affiliation{%
 \institution{National University of Singapore}
    \streetaddress{}
  \city{}
  \country{}
 }

\renewcommand{\shortauthors}{Li, et al.}

\begin{abstract}
With the information explosion on the Web, search and recommendation are foundational infrastructures to satisfying users' information needs. As the two sides of the same coin, both revolve around the same core research problem, matching queries with documents or users with items. In the recent few decades, search and recommendation have experienced synchronous technological paradigm shifts, including machine learning-based and deep learning-based paradigms. Recently, the superintelligent generative large language models have sparked a new paradigm in search and recommendation, \textit{i.e.}, generative search (retrieval) and recommendation, which aims to address the matching problem in a generative manner. In this paper, we provide a comprehensive survey of the emerging paradigm in information systems and summarize the developments in generative search and recommendation from a unified perspective. Rather than simply categorizing existing works, we abstract a unified framework for the generative paradigm and break down the existing works into different stages within this framework to highlight the strengths and weaknesses. And then, we distinguish generative search and recommendation with their unique challenges, identify open problems and future directions, and envision the next information-seeking paradigm.
\end{abstract}




\maketitle
\section{Introduction}
With the information explosion on the Web, a fundamental problem in information science gains increasing significance: sifting relevant information from vast pools to meet user needs. 
Nowadays, two information access modes --- search and recommendation --- serve as foundational infrastructures for information delivery. 
The objective of search is to retrieve a list of documents (\eg Web documents, Twitter posts, and answers) given the user's explicit query~\cite{kobayashi2000information}. By contrast, recommendation systems aim to recommend the items (\eg E-commerce products, micro-videos, and news) by implicitly inferring user interest from the user's profile and historical interactions~\cite{adomavicius2005toward}. 
Currently, search and recommendation systems are extensively employed across diverse scenarios and domains, including E-commerce, social media, healthcare, and education. 


Search and recommendation are the two sides of the same coin~\cite{belkin1992information}. 
Search is users' active information retrieval with explicit queries while recommendation is passive information filtering for users. 
Despite the distinct objectives, search and recommendation can be unified as the ``matching'' problem from a technical perspective~\cite{garcia2011information}. 
Search can be formulated as a matching between queries and documents, and recommendation can be considered a matching between users and items. 
The common matching property has spurred synchronous technological paradigm shifts in both search and recommendation systems. Specifically,
\begin{itemize}[leftmargin=*]
    \item \textbf{Machine learning-based search and recommendation.} 
    Related works, generalized ``learning to match''~\cite{xu2020deep}, learn a matching function using machine learning techniques (\eg learning to ranks~\cite{li2022learning, liu2009learning} and Matrix Factorization~\cite{lee2000algorithms}) to estimate the relevance scores on query-document pairs or user-item pairs.
    
    \item \textbf{Deep learning-based search and recommendation.} With the significant advancements in various neural networks such as CNN~\cite{lecun1998gradient}, RNN~\cite{hochreiter1997long}, GNN~\cite{scarselli2008graph}, and Transformer~\cite{vaswani2017attention}, search and recommendation have transitioned into the deep learning-based paradigm. This paradigm leverages the powerful representation ability of deep learning-based methods to encode inputs (\ie queries, documents, users, and items) into dense vectors in a latent space~\cite{karpukhin2020dense} and learn the non-linear matching functions. 
    
    \item \textbf{Generative search and recommendation.} 
    With the recent surge of generative large language models (LLMs), a new paradigm emerges in search and recommendation: generative search (retrieval)\footnote{Generative retrieval is another frequently used term.} and recommendation. Distinguished from previous discriminative matching paradigms, generative search and recommendation aim to directly generate the target document or item to satisfy users' information needs. 
 
\end{itemize}
Embracing generative search and recommendation brings new benefits and opportunities for the field of search and recommendation.
In particular, 
1) LLMs inherently possess formidable capabilities, such as vast knowledge, semantic understanding, interactive skills, and instruction following. These inherent abilities can be transferred or directly applied to search and recommendation, thereby enhancing information retrieval tasks. 
2)  The tremendous success of LLMs stems from their generative learning. A profound consideration to apply generative learning to search and recommendation, fundamentally revolutionizing the methods of information retrieval rather than only utilization of LLMs. 
3) LLMs-based generative AI applications, such as ChatGPT, are gradually becoming a new gateway for users to access Web content. Developing generative search and recommendation could be better integrated into these generative AI applications.

In this survey, we aim to provide a comprehensive review of generative search and recommendation from a technical perspective. 
We will not simply list and classify the current works, but instead abstract a unified framework for generative search and recommendation. Within the unified framework, we highlight the objectives of each stage and categorize existing works into different stages from a technical viewpoint. From the unified perspective, we could better highlight their commonalities and present their developments. 
In addition, we will delve into deeper discussions, including contrasting generative search and recommendation, identifying open problems of the generative paradigm, and envisioning future information-seeking paradigms. 

There are several key points that set our survey apart from existing ones. First and most important, we are the first to summarize and analyze the developments in generation search. Second, compared with existing generation recommendation surveys~\cite{deldjoo2024review,vats2024exploring,wu2023survey,lin2023can}, we focus on the fundamental generative paradigms for recommendation rather than incorporating  LLM-enhanced discriminative methods. Besides, we examine existing works from a technical perspective, using an abstract framework to categorize generation recommendation works into different stages. Third, we highlight the close relationship between generative search and generation recommendation, where many works in the two areas mutually inform and promote each other. By taking a unified view and comparing the respective developments, we are able to identify potential research directions.

The survey is structured as follows: Chapter 2 summarizes the previous paradigms in search and recommendation. Chapter 3 provides an overview of the generative paradigm for search and recommendation. In Chapters 4 and 5, we delve into the specific details of the generative paradigm in search and recommendation, respectively. Chapter 6 focuses on deep discussions, including the comparison of generative search and recommendation, open problems, and the next information-seeking paradigm. Finally, in Chapter 7, we give a conclusion of this survey.

\section{Traditional Paradigms}
\subsection{Machine Learning based Search and Recommendation}
\noindent$\bullet\quad$\textbf{Machine learning-based search}. 
In the era of machine learning, the core problem for search is to learn an effective function to predict the relevance scores between queries and documents. A series of classical works, such as
Regularized Matching in Latent Space (RMLS)~\cite{wu2013learning} and
Supervised Semantic Indexing (SSI)~\cite{bai2009supervised,bai2010learning}, explored mapping functions to transform features from queries and documents into a ``latent space''. The series of ``learning to rank'' algorithms~\cite{li2022learning, liu2009learning} were also proposed to develop effective rank losses for machine learning based search methods: the pointwise approaches~\cite{chu2005new, shashua2002ranking, nallapati2004discriminative} transform ranking into regression or classification on single documents; the pairwise approaches~\cite{burges2005learning, tsai2007frank, qin2007ranking} regard the ranking into classification on the pairs of documents; the listwise approaches~\cite{cao2007learning, qin2008query, wang2018lambdaloss, xia2008listwise, freund2003efficient} aim to model the ranking problem in a straightforward
fashion and overcome the drawbacks of the aforementioned two approaches by tackling the ranking problem directly.

\noindent$\bullet\quad$\textbf{Machine learning-based recommendation}. 
In the context of recommender systems, the user-item matching typically leverages Collaborative Filtering (CF), which assumes users with similar interactions (\eg ratings or clicks) share similar preferences on items~\cite{He2017Neural}. 
To achieve CF, early effort has been made to develop memory-based methods, which predict the user interactions by memorizing similar user's or item's ratings~\cite{breese1998empirical,herlocker1999algorithmic,linden2003amazon,sarwar2001item}. 
Later on, popularized by the Netflix Prize, Matrix Factorization (MF)~\cite{rendle2009bpr} has emerged as one of the most representative CF approaches. 
MF decomposes user-item interactions into latent factors for users and items in a latent space. It then predicts user-item interactions by matching these latent factors through inner product computations. 
Following MF, other notable methods have been proposed that also perform matching in the latent space, such as BMF~\cite{koren2009matrix} and FISM~\cite{kabbur2013fism}. 
In addition to the CF-based methods, an orthogonal line of research focuses on content-based techniques, aiming to encode user/item features for user-item matching. 
Factorization Machine~\cite{rendle2010factorization} is a prominent content-based method that represents user and item features as latent factors and models their high-order interactions to match between users and items.

\subsection{Deep Learning based Search and Recommendation}

\noindent$\bullet\quad$\textbf{Deep learning-based Search}. 
Deep learning-based search mainly relies on various neural architectures to represent queries and documents effectively. Feedforward neural networks are the first used to create semantic representations of queries and documents. Deep Structured Semantic Models were proposed to represent queries and documents with deep neural networks~\cite{huang2013learning}. The popular Convolutional Neural Networks are also explored for the application in capturing semantic embeddings~\cite{shen2014latent, hu2014convolutional,huang2017multi,socher2013reasoning}. Given the fact that both queries and documents are sequential texts, it is natural
to apply Recurrent Neural Networks to represent the queries and documents~\cite{palangi2016deep}. Later, with the rise of the Transformer architecture~\cite{vaswani2017attention} and pretrained BERT model~\cite{devlin2018bert}, the Bert-based dense retrievers show advanced performance in large-scale scenarios~\cite{karpukhin2020dense,qu2020rocketqa}. 

\noindent$\bullet\quad$\textbf{Deep learning-based recommendation}. 
As deep neural networks have demonstrated exceptional learning capabilities across various domains, there emerges a trend in leveraging deep learning methodologies to tackle complex interaction patterns for user-item matching in recommendation systems~\cite{kang2018self,chen2019dynamic}. 
Deep learning-based user-item matching can be broadly classified into two research directions. 
One research line focuses on matching function learning, which utilizes deep learning techniques to learn intricate user-item matching function~\cite{zhang2017joint}. 
Notably, Neural Collaborative Filtering (NCF)~\cite{He2017Neural} leverages Multi-Layer Perceptrons (MLP) to achieve expressive and complicated matching functions. 
This approach effectively models noisy implicit feedback data and enhances the recommendation performance. 
Another line of work lies in representation learning~\cite{xue2017deep}, which harnesses neural networks to transform user and item features into latent space conducive to matching. 
For instance, Bert4Rec~\cite{sun2019bert4rec} employs deep bidirectional self-attention mechanisms to transform user's historical sequences into latent space for sequential recommendation tasks. 
Caser~\cite{tang2018personalized} proposes a convolutional sequence model, which leverages both horizontal and vertical convolutional filters to identify complex historical interaction sequences.  
Subsequently, motivated by the graphical structure inherent in user-item interactions, researchers have explored the application of graph neural networks for recommendation tasks. 
This research direction has enabled the utilization of high-order neighbor information to enhance the representation of users and items for the user-item matching, as exemplified by NGCF~\cite{wang2019NGCF} and LightGCN~\cite{he2020lightgcn}.

\begin{figure*}[t]
\setlength{\abovecaptionskip}{0.1cm}
\setlength{\belowcaptionskip}{0cm}
\centering
\includegraphics[scale=0.82]{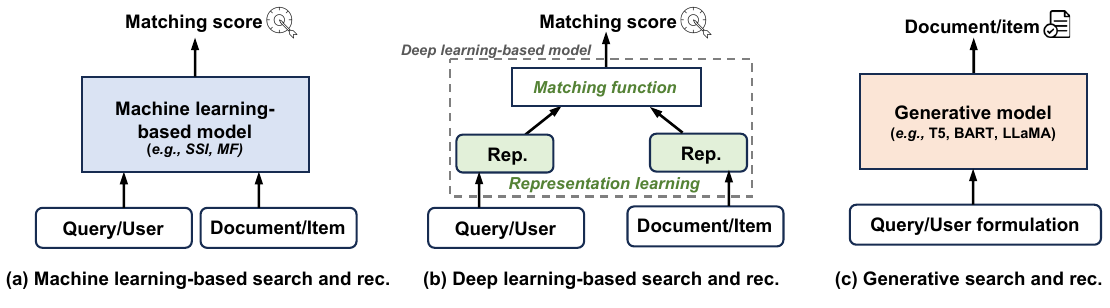}
\caption{Comparison of the three paradigms for search and recommendation, \ie machine learning-based, deep learning-based, and generative search and recommendation. ``rec.'' denotes ``recommendation''.}
\label{fig:paradigm}
\end{figure*}

\section{Generative Paradigm for Search and Recommendation}
In this section, we first clarify the scope of generative search and recommendation, including comparison with previous paradigms and comparison with other LLM-based methods. And then we abstract a unified framework for generative search and recommendation.

\subsection{Scope Clarification}
\textbf{Comparison with previous paradigms}. As depicted in Figure~\ref{fig:paradigm}, we present a summary of the three distinct paradigms in search and recommendation. Each paradigm employs different techniques to achieve the same goal of providing relevant documents/items for a given query/user. The machine learning and deep learning paradigms approach the search/recommendation task as a discriminative problem, focusing on calculating similarities between queries/users and documents/items. In contrast, the generative paradigm formulates the task as a generation problem, aiming to directly generate the documents/items based on the queries/users.

We need to address two specific issues for the three paradigms. 1) Deep learning is a subset of machine learning, and neural generative models are a subset of deep learning. While there are sub-concepts within these paradigms, they have clear boundaries and distinct development directions. For instance, with the emergence of deep learning, new methods have been developed to apply neural networks for feature extraction in search and recommendation systems, despite deep learning being a part of machine learning. Similarly, generative models have gained popularity, with numerous works focusing on generative search and recommendation. 2) We use the term ``paradigm shift'' to indicate potential community interest rather than actual trends. Both generative search and recommendation are relatively new approaches in the research community, and their effectiveness has not been fully verified historically.

\textbf{Comparison with LLM-based discriminative methods}. In this study, we define generative search and generative recommendation as methods that fully accomplish search and recommendation tasks using generative models. This is a specific definition that may exclude certain works. For example, some approaches~\cite{ma2023fine,muennighoff2024generative} may use generative language models to extract features from queries/users and documents/items. While these works also utilize generative language models, the overall paradigm is not significantly different from previous methods, as they simply replace the original encoder with a generative language model. In our study, we focus on presenting works that completely utilize generative paradigms to accomplish search or recommendation tasks.

\subsection{Unified Framework}
Benefiting from the simplicity of the generative paradigm, we could summarize the generative search and generative recommendation into a unified framework.

To accomplish search and recommendation tasks in a generative manner, there are four essential steps:
1) \textbf{Query/User Formulation}: This step aims to determine the input of the generative model. For search, complex query formulation is not necessary; for recommendation, user formulation is vital to transform user information into textual sequences.
2) \textbf{Document/Item Identifiers}: In practice, directly generating the document or item is almost impossible. Therefore, a short text sequence, known as the identifier, is used to represent the document or item.
3) \textbf{Training}: Once the input (query/user formulation) and output (document/item identifiers) of the generative model are determined, training is easily achieved via the generation loss.
4) \textbf{Inference}: After the training, the generative model can receive the query/user to predict the document/item identifier, and the document/item identifier can correspond to the document or item.

While the entire process may seem simple, achieving effective generative search and recommendation is not trivial. Numerous details need to be considered and balanced within the four steps mentioned above. In Sections 4 and 5, we will summarize generative search and generative recommendation methods, emphasizing their focus on specific aspects within the framework.

\section{Generative Search}
\subsection{Overview}
Generative search endeavors to leverage generative models, specifically generative language models, to complete the conventional search and retrieval process. The target is still the matching between documents and a given query. Different from previous paradigms, generative search aims to directly generate the desired target document when presented with a query.

\subsection{Query Formulation}
In search, users typically express their information needs through textual queries. This is in contrast to generative recommendation, which involves a ``user formulation'' step to convert user history into a textual sequence. In most retrieval tasks, the textual query can be directly inputted into the generative language model, sometimes with simple prefixes like ``query: ''. However, in certain retrieval tasks such as conversational QA and multi-hop QA, the query must be combined with the conversation context~\cite{li2023generative} or previous-hop answers~\cite{lee2022generative}.

\subsection{Document Identifiers}\label{sec:document_identifiers}
Ideally, generative search aims to directly generate the complete target document in response to a given query. However, in practice, this task proves to be extremely challenging for LLMs due to the length and inclusion of irrelevant information in the document's content. Consequently, current generative search approaches often rely on the use of identifiers to represent documents. These identifiers are concise strings that effectively capture the essence of a document's content. We summarize the current identifiers in generative search methods, and analyze their advantages and disadvantages as follows.

\textbf{Numeric ID}~\cite{tay2022transformer,wang2022neural,zhuang2022bridging,zhou2022dynamicretriever,nadeem2022codedsi,li2024towards}. Each document in the corpus can be assigned a unique numeric ID, such as ``12138''. During the inference stage, the LLM can receive the query as input and generate either a single numeric ID or a list of numeric IDs using beam search. Since each document corresponds to a numeric ID, the predicted numeric ID can represent a retrieved document. However, it is important to note that the numeric ID itself does not have any semantic relation to the content of the document. There are several methods to establish correlations between semantic text and numeric IDs. 1) Document-to-numeric ID training. During the training phase, an LLM can be trained to take a document's content as input and generate the corresponding numeric ID as the target. This training approach compels the LLM to memorize the relationships between documents and their numeric IDs. Once trained effectively, the LLM is expected to recall the numeric IDs of the documents accurately. 2) Clustered numeric IDs. In ~\cite{tay2022transformer}, the authors explored the concept of clustering document embeddings and assigning numeric IDs based on the cluster results. This approach allows similar documents with comparable content to be grouped into the same clusters, resulting in similar numeric IDs for these documents. Unlike randomly assigned numeric IDs, the clustered IDs are determined based on the content of the documents and are consistent with their semantics.

Numeric IDs pose challenges in generative search for the following reasons. 1) Generalization. The inability to generalize is a significant issue for numeric IDs. Previous studies~\cite{tay2022transformer} have shown that the Language Model can effectively memorize numeric IDs of passages in the training set. However, when it comes to the test set, the LLM's performance deteriorates. This can be attributed to the fact that numeric IDs lack semantic meaning, making it difficult for the model to generalize to unseen data. To address this problem, NCI~\cite{wang2022neural} proposed the inclusion of pseudo-queries on test set passages, which helps alleviate the issue. 2) Corpus update. Updating poses a challenge for generative search methods based on numeric IDs. Unlike dense retrieval methods that can easily update passages by modifying the embedding vectors in the corpus, the LLM struggles with updating passages. This is because the LLM stores the passages in its parameters, and accurately editing these parameters is not feasible. In~\cite{mehta2022dsi++}, the authors introduced incremental learning to solve the passage-adding problem partially. 3) Large-scale corpus. Since generative search models must memorize the associations between documents and their numeric IDs, the memorization difficulty increases as the document corpus sizes increase.  The work~\cite{pradeep2023does} explored to scaling up numeric ID based generative search. It is found that while generative search is competitive with state-of-the-art dual encoders on small corpora, scaling to millions of passages remains an important and unsolved challenge.

\begin{table*}
    \centering
    \caption{Identifiers in generative search. }
    \begin{tabular}{|c|c|c|c|c|c|}\hline
         Type&  Semantic& Distinctiveness& Update&Training& Applicable retrieval scenarios\\ \hline
         Numeric ID& $\times$ & $\checkmark$& $\times$ &\makecell[c]{document-to-identifier \\ query-to-identifier}& Small-scale corpus\\ \hline
         Titles&$\checkmark$&$\times$& $\checkmark$ &query-to-identifier&Document-level retrieval\\\hline
         N-grams&$\checkmark$&$\times$& $\checkmark$ &query-to-identifier&Document-level \& Passage-level retrieval \\\hline
         Codebook&$\checkmark$&$\checkmark$& $\checkmark$ &\makecell[c]{codebook training \\ query-to-identifier}&Document-level \& Passage-level retrieval \\\hline
         Multiview&$\checkmark$&$\times$& $\checkmark$ &query-to-identifier&Document-level \& Passage-level retrieval \\\hline
    \end{tabular}
    \label{tab:identifiers_in_search}
\end{table*}

\textbf{Document titles}~\cite{de2020autoregressive,li2023generative,lee2022generative,chen2022gere}. In certain specific scenarios, documents possess titles that can serve as effective identifiers. For instance, in Wikipedia, each page is assigned a unique title that succinctly summarizes its content. These titles are semantically linked to the documents and establish a one-to-one correspondence, making them ideal identifiers. In 2021, Cao et al. investigated the utilization of titles as identifiers in entity retrieval and document retrieval~\cite{de2020autoregressive}. Besides, there are some title-like identifiers, including URLs~\cite{ren2023tome, zhou2022ultron, ziems2023large}, keywords~\cite{lee2023glen}, and summaries~\cite{zhou2023enhancing}.

However, document titles are only effective in certain retrieval domains and prove ineffective in the following aspects. 1) Firstly, in passage-level retrieval, documents are often divided into smaller retrieval units known as passages. It is hard to design effective passage identifiers even based on the document titles, which makes title-based generative search perform badly in passage retrieval. In 2023, Li et al. employed document titles plus section titles as passage identifiers, but this approach was limited to the Wikipedia corpus~\cite{li2023generative}. 2) Secondly, web pages differ from Wikipedia in web retrieval as they lack high-quality titles. These titles may not accurately represent the content of the document, and multiple web pages may share the same title. 3) Additionally, numerous pages do not have any titles at all. These factors contribute to generative search lagging behind traditional retrieval methods.

\textbf{N-grams}~\cite{bevilacqua2022autoregressive, zhang2023term,chen2023unified,wang2023novo}. The document itself is semantic but cannot serve as a reliable identifier. This is because not all of its content is necessarily related to the user query, making it difficult for the LLM to generate irrelevant content based solely on the user query. However, inspired by this limitation, the N-grams within the document that are semantically related to the user query could be regarded as potential identifiers. In 2022, the authors in ~\cite{bevilacqua2022autoregressive} trained an LLM to generate target N-grams using a user query as input. These N-grams were selected based on the word overlap between the user query and the N-grams. Once trained, the LLM was able to predict relevant N-grams given a query. These predicted N-grams were then transformed into a passage rank list using a heuristic function. The proposed method was evaluated on commonly used datasets, including NQ and TriviaQA, rather than some specifically designed subsets of datasets.

However, N-gram identifiers have certain limitations. 1). Firstly, they are not as discriminative as numeric IDs. Unlike numeric IDs, N-grams cannot directly correspond to specific documents on a one-to-one basis. This means that the LLM  must rely on a transformation function to convert the predicted N-grams into a document ranking list. In SEAL~\cite{bevilacqua2022autoregressive}, for example, the transformation function calculates a document's score by summarizing the scores of the N-grams it contains. In a way, this transformation function acts as a simple retrieval approach, preventing N-gram-based generative search from achieving end-to-end document retrieval. 2) Secondly, the selection of N-grams in the training phase is a crucial aspect. A document may contain numerous N-grams, and it is necessary to select those that are semantically relevant to the query. However, the number of N-grams to be selected and the method used for selection are adjustable and can vary largely.

\textbf{Codebook}~\cite{sun2023learning,zeng2023scalable,yang2023auto,zhang2024model}. The text codebook, also known as tokens, plays a fundamental role in LLMs. Text is inherently discrete, allowing LLMs to acquire knowledge through tasks like predicting the next token. Some studies ~\cite{van2017neural, lee2022autoregressive} have also focused on developing visual codebooks for discrete images. As mentioned earlier, while a document can be represented as a list of tokens, it is not suitable as an identifier due to its lengthiness. Therefore, an alternative approach is to learn a new codebook specifically for documents, which can represent them more efficiently than natural text tokens. In 2023, Sun et al. proposed a method to learn a codebook for documents in generative search~\cite{sun2023learning}.  The work~\cite{yang2023auto} proposed an end-to-end framework to automatically search best identifiers according to the document's content. 

However, learning a codebook for documents is a complex process. It typically involves encoding documents into dense vectors using an encoder network, discretizing these vectors, and then using a decoder network to reconstruct the original document. The size of the codebook and the length of the sequence required to represent a document must be carefully adjusted. Additionally, compared to titles and n-grams, the codebook lacks interpretability.

\begin{table*}
    \centering
    \caption{Summary of the representative generative search methods.}
    \begin{tabular}{|c|c|c|c|c|}\hline
     Method&  Identifier& \makecell[c]{Backbone}&\makecell[c]{Constrained \\generation}& Datasets\\ \hline
     GENRE~\cite{de2020autoregressive}& Titles & BART-large&Trie& KILT\\ \hline
     DSI~\cite{tay2022transformer}& Numeric ID & \makecell[c]{T5-base, T5-large, \\T5-XL, T5-XXL} &Trie& \textit{NQ10K}, \textit{NQ100K}, \textit{NQ320K} \\ \hline
     SEAL~\cite{bevilacqua2022autoregressive}& N-grams& BART-large &FM-index& KILT, \textit{NQ10K}, NQ \\ \hline
     NCI~\cite{wang2022neural}& Numeric ID&  \makecell[c]{T5-small, T5-base, \\T5-large} &Trie& \textit{NQ320K}, \textit{Trivia QA (subset)}\\ \hline
     MINDER~\cite{li2023multiview}& Multiview&  BART-large &FM-index& NQ, TriviaQA, MSMARCO\\ \hline
     GENRET~\cite{sun2023learning}& Codebook&  T5-base &Trie& \textit{NQ320K}, \textit{MSMARCO}, BEIR\\ \hline
     TOME~\cite{ren2023tome}& \makecell[c]{Title-like (URLs)}&  T5-large, T5-XL &Trie& \textit{NQ320K}, MSMARCO\\ \hline
     LTRGR~\cite{li2023learning}& Multiview&  BART-large &FM-index& NQ, TriviaQA, MSMARCO\\ \hline
     DGR~\cite{li2024distillation}& Multiview&  BART-large &FM-index& NQ, TriviaQA, MSMARCO, TREC DL\\ \hline
     DSI(scaling up)~\cite{pradeep2023does}& Numeric ID&  \makecell[c]{T5-base, T5-large, \\T5-XL, T5-XXL }&Trie&\textit{NQ320K}, \textit{Trivia QA (subset)}, MSMARCO\\ \hline
     DSI-QG~\cite{zhuang2022bridging}& Numeric ID&  T5-base, T5-large&-&\textit{NQ320K}, \textit{XOR QA 100k} \\ \hline
    DSI++~\cite{mehta2022dsi++}& Numeric ID&  \makecell[c]{T5-base, T5-large, \\T5-XL}&Trie&\textit{NQ-inc}, \textit{MSMARCO-inc}\\ \hline
    GCoQA~\cite{li2023generative}& Titles&  \makecell[c]{T5-small, T5-base, \\T5-large, T5-XL}&Trie&\textit{TOPIOCQA}, \textit{QRECC},\textit{ORQUAC}\\ \hline
     GMR~\cite{lee2022generative}& Titles&  T5-base&Trie&HotpotQA\\ \hline
     CorpusBrain~\cite{chen2022corpusbrain}& Titles& BART-large&Trie&KILT\\ \hline
     UGR~\cite{chen2023unified}& N-grams& BART-large&FM-index&KILT\\ \hline
     SE-DSI~\cite{tang2023semantic}& \makecell[c]{Title-like \\(pseudo-queries)}& T5-base&Trie&\textit{NQ100K}, \textit{MSMARCO}\\ \hline
     GenRRL~\cite{zhou2023enhancing}& \makecell[c]{Title-like (summaries)}& T5-base&Trie&\textit{NQ320K}, \textit{MSMARCO}\\ \hline
     GLEN~\cite{lee2023glen}& \makecell[c]{Title-like (keywords)}& T5-base&Trie&\textit{NQ320K}, MSMARCO, BEIR\\ \hline
     Ultron~\cite{zhou2022ultron}& \makecell[c]{Title-like (URLs)}& T5-base&Trie&\textit{NQ320K}, \textit{MSMARCO}\\ \hline
     LLM-URL~\cite{ziems2023large}& \makecell[c]{Title-like (URLs)}& GPT3&-&NQ, TriviaQA, WebQuestions\\ \hline
     ASI~\cite{yang2023auto}& \makecell[c]{Codebook \\(Auto Search)}& T5-like&-&MSMARCO (3.1M)\\ \hline
     RIPOR~\cite{zeng2023scalable}& \makecell[c]{\makecell[c]{Codebook \\(Residual  Quantization)}}& T5-base&Trie&MSMARCO, TREC DL\\ \hline
     ListGR~\cite{tang2024listwise}& \makecell[c]{\makecell[c]{Numeric ID}}& T5-base&Trie&\textit{MSMARCO 100K}, \textit{NQ320K}\\ \hline
     MEVI~\cite{zhang2024model}& \makecell[c]{\makecell[c]{Codebook}}& T5-large&Trie&MSMARCO, NQ\\ \hline
    \end{tabular}
    \label{tab:generative_retrieval_methods}
\end{table*}

\textbf{Multiview identifiers}~\cite{li2023multiview,li2023learning,li2024distillation}. The above identifiers are limited in different aspects: numeric IDs require extra memorization steps and are ineffective in the large-scale corpus, while titles and substrings are only pieces of passages and thus lack contextualized information. More importantly, a passage should answer potential queries from different views, but one type of identifier only represents a passage from one perspective. Therefore, a natural idea is to combine different identifiers to exploit their advantages. In 2023, Li et al.~\cite{li2023multiview} proposed a generative search framework, MINDER, to unify different identities, including titles, N-grams, pseudo queries, and numeric IDs. Any other identifiers could also be included in this framework. The experiments on three common datasets verify the effectiveness and robustness of different retrieval domains benefiting from the multiview identifiers. 

Similar to the N-grams, multiview identifiers cannot correspond to documents one-to-one and thus require the transformation function. Besides, in the inference stage, the LLM needs to generate different types of identifiers, decreasing the inference efficiency.

\textbf{Document identifier summary}. To better illustrate the characteristics of different document identifiers, we summarize Table~\ref{tab:identifiers_in_search} from the aspects of semantic, distinctiveness, update, training, and applicable retrieval domains.  Among the various types of identifiers, the codebook shows great potential in all dimensions. However, it does have a drawback in that it requires a complex training process. The multiview identifier has a simpler training process, but it lacks distinctiveness and requires a transform function from identifiers to documents.
\subsection{Training}
In contrast to traditional retrieval methods, the training for generative search is notably simpler. We categorize the training into generative and discriminative training two categories.

\begin{figure*}[t]
\setlength{\abovecaptionskip}{0cm}
\setlength{\belowcaptionskip}{-0cm}
\centering
\includegraphics[scale=0.52]{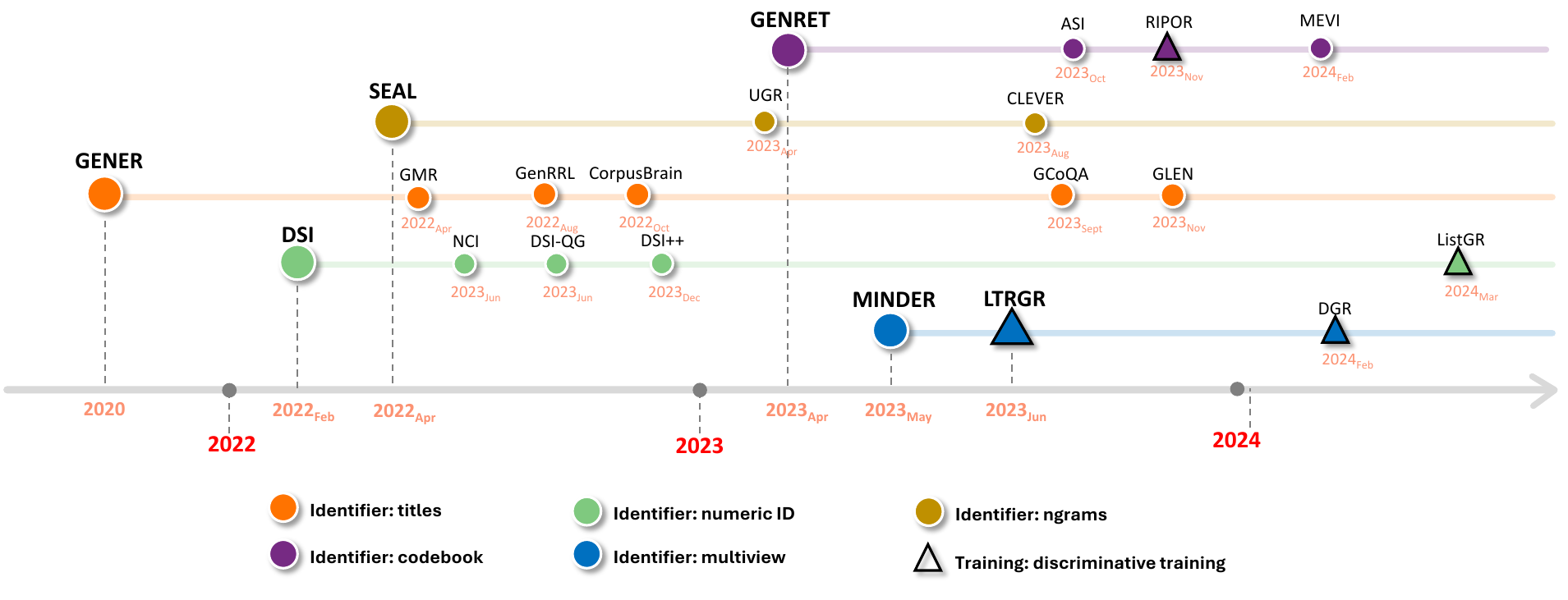}
\caption{{Major milestones in the development of generative search.}}
\label{fig:timeline_search}
\end{figure*}

\textbf{Generative training}. Once the input and output are determined, the LLM can be trained to predict the next token. There are two main training directions. 1) \textbf{Query-to-identifier} training. This involves training the LLM to generate the corresponding identifier for a given query. Most identifier types require this training, with some, such as document titles, N-grams, and multiview identifiers, only needing this query-to-identifier training direction. 2) \textbf{Document-to-identifier} training. In this approach, the LLM learns to predict the corresponding identifier when given a document as input. This training is crucial for certain identifiers, such as numeric IDs and the codebook, as they need to align with the semantics of documents, and \textbf{codebook training} could be regarded as the special document-to-identifier training. It is noted that not all documents in a search have labels (queries), making it challenging for the LLM to memorize these documents. To address this, some generative search methods~\cite{zhuang2022bridging, pradeep2023does} utilize pseudo pairs to expand the training samples and enhance document memorization.

\textbf{Discriminative training}. Generative search presents a new paradigm for retrieval, as it transforms the original discriminative methods into the generative methods and could train the retrieval model (generative language model) via generation loss. However, the work~\cite{li2023learning} has highlighted the importance of discriminative training in generative search. It has been observed that discriminative training can further improve a well-trained generative model. This finding is meaningful for both generative search and the traditional retrieval paradigm. Previous retrieval studies have developed numerous discriminative losses (rank losses) and negative sample mining methods, which is a big treasure for the retrieval field. As the finding illustrates, the previous research approaches could be adjusted to enhance the current generative search. The following works~\cite{zeng2023scalable, tang2024listwise} further verified the effectiveness of introducing discriminative training in generative search.

\subsection{Inference}
After completing the training process, the generative search model can be used for retrieval purposes.

\textbf{Free generation}. During the inference stage, the trained LLM is able to predict identifiers based on a user query, similar to the training process. This generation process is free, as the LLM could generate any text without any constraints. These predicted identifiers may correspond directly to specific documents, or they may be determined through a heuristic function based on the type of identifier. This is the unique aspect of generative search, as it allows for direct document retrieval through generation. However, in practical applications, only a few generative approaches~\cite{yang2023auto, ziems2023large} actually utilize the free generation method. Because the scope of identifiers is limited, the potential for generation is infinite. This means that the LLM may generate identifiers that could not belong to any document within the corpus.

\textbf{Constrained generation}. Most generative search approaches employ constrained generation to guarantee the LLM generates valid identifiers. This technique involves post-processing to mask any invalid tokens and only allows the generation of valid tokens that belong to identifiers. To achieve this, special data structures such as Trie and FM\_index~\cite{ferragina2000opportunistic} are utilized. The FM\_index enables the LLM to generate valid tokens from any position within the identifier, while the Trie only supports generation from the first token of an identifier. These data structures play a key role in enabling the LLM to accurately generate valid identifiers.

The predicted identifiers could correspond directly to specific documents, or they may be determined through a heuristic function based on the type of identifier. Finally, generative search methods could give a final document ranking list in a generative way.

\subsection{Summary}
\textbf{Methods summary}. We summarize the current generative search methods in Table~\ref{tab:generative_retrieval_methods}, from the aspects of identifier, backbone, constrained generation, and datasets. 1) Identifier. We have fully discussed the characteristics and problems of different identifiers in the previous sections and Table~\ref{tab:identifiers_in_search}. Different identifiers usually require different training strategies and inference processes. 2) Backbone. Almost all methods employ pretrained language models, like BART~\cite{lewis2019bart} and T5~\cite{raffel2020exploring}, as the backbone, due to their extensive language knowledge. However, it is also noted that few current approaches apply advanced large foundation models, like ChatGPT and LLaMA. On the one hand, increasing the model size alone may not lead to significant research contributions; on the other hand, closed-source models cannot be adjusted for constrained generation. 3) Constrained generation. Almost all methods adopt the constrained generation to guarantee the valid identifier generation. The multiview identifier and N-gram identifier require the FM\_index structure, while others need the Trie structure to facilitate the constrained generation.  4) Datasets are primarily focused on document-level and passage-level retrieval tasks, with some also designed for conversational QA, multi-hop QA, and cross-lingual retrieval tasks. However, due to limitations in identifier types, some methods have had to reformulate certain datasets for evaluation, such as NQ320k, TriviaQA (subset), and MSMARCO (subset). While these reformulated datasets may highlight the advantages of generative search, they may not align with real-world applications. Multiview identifiers based methods~\cite{li2023multiview,li2023learning, li2024distillation} have a significant advantage in general retrieval datasets.

\textbf{Timeline summary}. We also provide a brief overview of the development of generative search based on the timeline, as depicted in Figure~\ref{fig:timeline_search}. We specifically highlight the works that first introduced a new identifier type or training scheme in generative search. The credit for the first generative search work goes to GENRE~\cite{de2020autoregressive}. While GENRE's primary focus is on entity retrieval rather than document retrieval, it was the first to employ the autoregressive generation paradigm to accomplish the retrieval task and introduce constrained beam search. Starting from 2022, there has been a continuous introduction of new identifier types~\cite{tay2022transformer,bevilacqua2022autoregressive,sun2023learning,li2023multiview}. In 2023, discriminative training~\cite{li2023learning} was introduced in generative search, followed by the latest works~\cite{zeng2023scalable,tang2024listwise}.

\subsection{LLMs for Retrieval beyond Generative Search}
There have been efforts to investigate the potential applications of LLMs in text retrieval beyond generative search. These studies have concentrated on leveraging LLMs to improve existing retrieval pipelines by substituting smaller language models with larger ones.  1) \textbf{LLMs for query expansion}. Some works~\cite{wang2023query2doc, jagerman2023query} utilized LLMs to expand the queries. For instance, Query2Doc~\cite{wang2023query2doc} employs LLMs to generate synthetic documents, which can then be integrated into traditional document retrieval systems to enhance their performance. 2) \textbf{LLMs as feature extractors}. Dense retrievers are usually conducted based on encoder-only language models, like BERT. Recently, some work~\cite{ma2023fine,muennighoff2024generative} has focused on leveraging generative large language models to improve document representations. 3) \textbf{LLMs based rankers}. Rankers should further refine the order of the retrieved candidates to improve output quality, and LLMs have also been applied to refine the ranking of retrieved documents, as demonstrated in studies~\cite{sun2023chatgpt, qin2023large}.

\section{Generative Recommendation}
\subsection{Overview}\label{sec:rec_overview}
Since recommendation systems aim to filter out the items that are relevant to the user interests, two crucial components are required to achieve matching in the natural language space, \textit{i.e.}, user formulation, and item identifiers. 
The user formulation and item identifiers are analogous to the query formulation and document identifier in generative search, respectively. 
Specifically, as the input for generative model, the user formulation contains diverse information (\textit{e.g.}, user's historical interactions, and user profile) to represent a user in the natural language for modeling the user interests. 
To match the user interests in natural language, generative recommendation system will generate item identifier of the relevant items. 

\subsection{User Formulation}\label{sec:user_formulation}
Since there is no specific ``query'' for each user in recommendation systems, user formulation is a crucial step to represent a user for personalized recommendation. 
In generative recommendation systems, the user\footnote{Here, the ``user'' refers to the input of the generative recommendation task.} is primarily formulated based on four distinct components: the task description, user-associated information, context information, and external knowledge expressed in natural language. 
In particular, the user-associated information mainly consists of user's historical interactions and user profile (\textit{e.g.,} age and gender). 
Based on the availability of the data, previous work formulates the user by encompassing one or multiple components through pre-defined prompt templates for each user. 

\textbf{Task description}~\cite{geng2022recommendation}. 
To leverage the strong comprehension ability of the powerful generative models, task description is employed to guide the generative models to accomplish the recommendation task, \ie next-item prediction. 
For instance, in~\cite{bao2023bi}, the task description for movie recommendation is articulated as ``Given ten movies that the user watched recently, please recommend a new movie that the user likes to the user.''. 
It is noted that the task description can also serve as the prompt template, where various components are inserted into the template to formulate the user. 
For example,~\cite{geng2022recommendation} uses ``I find the purchase history of ..., I wonder what is the next item to recommend to the user. Can you help me decide?'' as the task description to instruct the generative models, where the user's historical interactions will be utilized for user formulation. 
In~\cite{zhang2023recommendation}, ``Here is the historical interactions of a user: ..., his preferences are as follows: ..., please provide recommendations.'' is used to guide the generative models, where historical interactions and user preference are incorporated to formulate the user. 

\textbf{User's historical interactions}~\cite{kim2024large,zhang2023collm}. 
Serving as the implicit user feedback on items, user's historical interactions play a crucial role in representing user behavior~\cite{he2020lightgcn} in recommender systems, which implicitly conveys user preference over items. 
To present the historical interactions, \textit{i.e.}, the sequence of interacted items, one common practice is to utilize the item ID to form the interaction sequence~\cite{geng2022recommendation,hua2023index}. 
Nevertheless, as pointed out by~\cite{yu2024ra}, generative models face limitations in capturing collaborative information while excelling in capturing nuanced item semantics. 
Therefore, two categories of work emerge to further improve the recommendation accuracy. 1)
To better leverage the strong semantics understanding of LLMs, a line of work~\cite{cui2022m6,chen2023palr,hou2023large,liu2024rec,kim2024large} attempts to integrate rich side information of items in historical interactions. 
In particular,~\cite{liu2023llmrec,cui2022m6} incorporates item descriptions when listing the user's historical interactions.~\cite{lin2023multi} utilizes item titles and item attributes to exploit the rich semantics of items for a better understanding of user preference. 
2) To strengthen the collaborative information understanding of LLMs, another line of work aims to incorporate ID embeddings of the interacted items for LLMs to understand user behavior. 
For example, LLaRA~\cite{liao2023llara} additionally includes ID embedding after the title's token embedding to represent each item in historical interactions. 
Moreover, with the advancements in multimodal LLMs, some work~\cite{liu2024rec,geng2023vip5} attempts to incorporate multimodal feature of the item, \textit{e.g.,} visual features, to complement the textual user's historical interactions. 


\textbf{User profile}~\cite{gao2023chat,lu2024aligning}. 
To enhance the user modeling, integrating user profile (\textit{e.g.,} demographic and preference information about the user) is an effective way to model the user characteristics in recommender systems~\cite{uyangoda2018user}. 
In most cases, user's demographic information (\textit{e.g.,} gender) can be obtained directly from the online recommendation platform. 
Such user information is then combined with descriptional text, \textit{e.g.,} ``User description: female, 25-34, and in sales/marketing''~\cite{xi2023towards}. 
For instance,~\cite{gao2023chat} utilizes user's age and gender to prompt ChatGPT to enhance its comprehension of user characteristics based on encoded prior knowledge. 
Although demographic information can implicitly reflect the general preference within specific user groups (\textit{e.g.,} teenagers), 
it can further enrich models' comprehension of the user by explicitly detailing user's general preference and current intention during user formulation. 
To obtain the explicit general preference and current intention,~\cite{zhang2023recommendation} proposes to leverage LLMs to infer the user intention and the general preference based on the user's historical interactions using tailored prompts. 
Besides,~\cite{zheng2024harnessing} uses LLMs to summarize user preference based on the user's historical interactions. 
However, acquiring user profiles can be challenging due to user privacy concerns, leading some studies to employ user ID to capture the collaborative information~\cite{geng2022recommendation} or discard user profile for user formulation~\cite{rajput2023recommender}. 

\textbf{Context information}~\cite{liu2024large,yu2024ra}.
In addition to user's historical interactions and profile, environmental context information (e.g., location and time), which can influence user decisions, is also advantageous for models to better match the users with appropriate items. 
For example, users may prefer to purchase a coat rather than a T-shirt in winter in clothing recommendation. 
Therefore, incorporating context information such as time, can achieve effective user understanding in real-world application scenarios. 
For instance,~\cite{liu2024large} harnesses diagnosis and procedures for medication recommendation while~\cite{yu2024ra} incorporates context information via learnable soft prompt to capture unobserved context signals.

\textbf{External knowledge}~\cite{du2024large}. 
Although generative models have demonstrated promising performance in recommendation tasks based on user-associated information, recent research has explored leveraging external knowledge to enhance the performance of generative recommender models. 
To harness structured information from user-item graph,~\cite{du2024large} integrates the graph data in natural language and further propagates higher-order neighbor's information to capture complex relations between users and items. 
Additionally,~\cite{luo2023recranker} leverages external knowledge from conventional recommender models by incorporating prediction from these models in natural language, showcasing collaborative efforts from both conventional models and generative models. 
Some work also incorporates an item candidate set to reduce the searching space of the whole item set, thus alleviating the hallucination problem and improving the accuracy~\cite{dai2023uncovering,zhang2023recommendation}. 


\begin{table*}[t]
\centering
\caption{Identifiers in generative recommendation.}
\begin{tabular}{|c|c|c|c|c|}
\hline
Type & Semantic & Distinctiveness  & Update & Training  \\ \hline
Numeric ID & $\times$ & $\checkmark$  & $\times$ & user-to-identifier \\ \hline
Item's textual metadata & $\checkmark$ & $\times$  & $\checkmark$ & user-to-identifier \\ \hline
Codebook & $\checkmark$ & $\checkmark$ & $\checkmark$ & item-to-identifier, user-to-identifier, auxiliary alignment \\ \hline
Multi-facet & $\checkmark$ & $\checkmark$ & $\checkmark$ & user-to-identifier \\ \hline
\end{tabular}
\label{tab:recommendation_identifier}
\end{table*}

\subsection{Item Identifiers}\label{sec:rec_identifier}
Similar to generative search, the generative recommender models are expected to generate relevant items given the user formulation. Nevertheless, items in recommendation platforms usually consist of various side information from different modalities, \textit{e.g.,} thumbnails of the micro-videos, audios of the music, and titles of the news. 
As such, the complex data from the items in recommendation necessitates item identifiers to present each item's characteristics in the language space for generative recommendation. 
It is highlighted that a good item identifier should at least meet two criteria as pointed out in~\cite{lin2023multi}: 
1) distinctiveness to emphasize the salient item features learned from user behaviors. And 
2) semantics to focus on the utilization of prior knowledge in pre-trained language models, which facilitates strong generalization to cold-start and cross-domain recommendation. 
Existing work usually constructs the item identifiers in the following four strategies, meeting different criteria accordingly. 

\textbf{Numeric ID}~\cite{geng2022recommendation,cao2024aligning,wang2024enhanced,si2023generative, chen2023continual,hua2023up5,petrov2023generative,zhai2023knowledge,li2023prompt,qiu2023controlrec}.
Given the widely demonstrated efficacy of numeric IDs for capturing collaborative information in traditional recommendation models~\cite{zhang2023collm}, a straightforward strategy in generative recommendation frameworks is to adopt the strategy of using numeric IDs to represent items~\cite{geng2022recommendation,li2024pap}. 
However, directly adopting the ID setting in traditional recommender models is infeasible in generative recommender models. 
Because traditional recommender models consider each item as an independent ``token'', which cannot be further tokenized and strictly refers to one independent embedding. 
This requires adding all the independent tokens into the model, which requires 1) large memory to store every item embedding, and 2) sufficient interactions for training the item embedding. 
To combat these issues, generative recommender models offer a promising solution by designing the item identifier as a token sequence, where the numeric IDs can be further tokenized and are associated with several token embeddings. 
As such, a token sequence as a numeric ID makes it possible to use finite tokens to represent infinite items~\cite{hua2023tutorial,li2023large}. 

To effectively represent an item with a numeric ID in a token sequence, previous work explores different strategies for ID assignment~\cite{hua2023index}. 
In~\cite{geng2022recommendation}, sequential indexing is utilized to capture the collaborative information in an intuitive way. 
Specifically, sequential indexing represents the user's items in chronological order by consecutive numeric IDs, (\textit{e.g.,} ``11138'', ``11139'', $\dots$, ``11405''), which could capture the co-occurrence of items that are interacted with the same user. 
However, such sequential indexing might suffer from the potential data leakage issue~\cite{hua2023index,rajput2023recommender}.~\cite{hua2023index} fixes this potential issue and explores other non-trivial indexing methods that incorporate prior information on the items, such as semantics and collaborative knowledge. 
By constructing IDs based on item categories in a hierarchical structure, the items belonging to the same categories will have similar IDs and are validated to be effective in generative recommendation. 
Similarly, SEATER~\cite{si2023generative} proposes a tree-based hierarchical identifier with numeric IDs, where items with similar interactions will have similar IDs. 
In addition,~\cite{hua2023index} also attempts to construct IDs based on the item co-occurance matrix, where items that co-occur more times will have more similar IDs, which is assessed to be beneficial in generating appropriate recommendation. 

Despite the effectiveness of distinctive numeric IDs in generative recommendation, it usually lacks semantic information, thus suffering from the cold-start problem~\cite{lin2023multi} and failing to leverage the world knowledge encoded in the recently emerged powerful generative models, \textit{e.g.,} LLMs. 



\textbf{Item's textual meta data}~\cite{bao2023bi,zhang2023chatgpt,lin2023sparks,harte2023leveraging,yin2023heterogeneous,di2023retrieval,luo2024integrating,liu2023recprompt,wang2023drdt,sun2023large,yue2023llamarec,wang2023multiple}.
To overcome the absence of semantics in numeric IDs, other work~\cite{bao2023bi,kim2024large} utilizes the item's textual metadata, \textit{e.g.,} item title, leveraging the LLMs' world knowledge encoded in the parameters to better comprehend the item characteristics based on the semantics in the item's textual descriptions. 
For instance,~\cite{bao2023bi,liao2023llara,zhang2021language,ji2023genrec,wang2023zero,hou2023large} use movie title,~\cite{liu2024rec,ji2023genrec,li2023gpt4rec,chen2023palr,liu2023chatgpt,zhang2023recommendation} use the product name,~\cite{zhiyuli2023bookgpt} utilizes book name,~\cite{li2023pbnr,li2023preliminary} adopt news title,~\cite{dai2023uncovering} uses song title,~\cite{tan2024towards} employs abstractive text of items, and~\cite{cui2022m6} includes both title and descriptions of online products, as the item identifier for recommendation. 
Although leveraging an item's textual metadata significantly alleviates the cold-start issue~\cite{bao2023bi,lin2023multi}, it is still suboptimal for effective recommendation. 
The textual metadata, especially the descriptions could be very lengthy, which would cause out-of-corpus issues, \textit{i.e.,} the generated token sequence could not match any valid item identifier. 
Although grounding the generated tokens to existing items via distance-based methods is a potential solution to address this problem~\cite{bao2023bi}, 
it would take us back to deep learning-based recommendation since we need to calculate the matching score between the generated item and each in-corpus item~\cite{li2023large}. 

\textbf{Codebook}~\cite{rajput2023recommender}. 
To simultaneously leverage the semantics while pursuing a unique short token sequence,~\cite{rajput2023recommender} proposes to learn a codebook to construct the item identifier in generative recommendation. 
Similar to the codebook for document identifier in generative search, related work in recommendation~\cite{rajput2023recommender,hou2023learning,hou2023learning} focuses on developing a codebook specifically for items. 
Typically, RQ-VAEs~\cite{zeghidour2021soundstream} are utilized to learn the codebook, where the input is the item's semantic representation extracted from a pre-trained language model (\textit{e.g.,} LLaMA~\cite{touvron2023llama}) and the output is the generated token sequence. 
The overall process of training codebook is similar to that in generative search, as discussed in Section~\ref{sec:document_identifiers}. 
Along this line, TIGER~\cite{rajput2023recommender} is a representative work that generates an item's semantic ID based on the item's textual descriptions via the codebook. 
LC-Rec~\cite{zheng2023adapting} further enhances the generated ID's representation to align with user preference and semantics of the item's textual descriptions. 

However, while codebook-based identifiers meet both semantics and distinctiveness, they suffer from the misalignment between the semantics correlation and interaction correlation. 
Specifically, the codebooks essentially capture the correlation of the item semantics into the identifier, \ie items with similar semantics will have similar identifiers. 
The identifier representations will then be optimized to capture interaction correlation by training on recommendation data. 
Nevertheless, the items with similar codes might not necessarily have similar interactions, thereby hurting the learning of user behavior. 

\begin{table*}[t]
\centering
\caption{Summary of the representative generative recommendation methods.}
\setlength{\tabcolsep}{1.2mm}{
\resizebox{\textwidth}{!}{
\begin{tabular}{|c|c|c|c|c|c|c|}
\hline
\textbf{Item identifier} & \textbf{User formulation} & \textbf{Method} & \textbf{Backbone} & \textbf{Generation} & \textbf{Dataset} & \textbf{Domain} \\ \hline
\multirow{15}{*}{Numeric ID} & \multirow{2}{*}{\begin{tabular}[c]{@{}l@{}}Task description\\ Historical interactions\end{tabular}} & \multirow{2}{*}{RecSysLLM~\cite{chu2023leveraging}} & \multirow{2}{*}{GLM-10B} & \multirow{2}{*}{Trie} & \multirow{2}{*}{Sports, Beauty, Toys} & \multirow{2}{*}{E-commerce} \\
 &  &  &  &  &  &  \\ \cline{2-7} 
 & \multirow{6}{*}{\begin{tabular}[c]{@{}l@{}}Task description\\ User profile\\ Historical interactions\end{tabular}} & \multirow{3}{*}{P5~\cite{geng2022recommendation}} & \multirow{3}{*}{T5-small, T5-Base} & \multirow{3}{*}{Free} & \multirow{3}{*}{\begin{tabular}[c]{@{}c@{}}Sports, Beauty, Toys\\ Yelp\end{tabular}} & \multirow{3}{*}{\begin{tabular}[c]{@{}c@{}}E-commerce\\ Restaurant\end{tabular}} \\
 &  &  &  &  &  &  \\
 &  &  &  &  &  &  \\ \cline{3-7} 
 &  & \multirow{3}{*}{How2index~\cite{hua2023index}} & \multirow{3}{*}{T5-small} & \multirow{3}{*}{Trie} & \multirow{3}{*}{\begin{tabular}[c]{@{}c@{}}Sports, Beauty\\ Yelp\end{tabular}} & \multirow{3}{*}{\begin{tabular}[c]{@{}c@{}}E-commerce\\ Restaurant\end{tabular}} \\
 &  &  &  &  &  &  \\
 &  &  &  &  &  &  \\ \cline{2-7} 
 & \multirow{3}{*}{\begin{tabular}[c]{@{}l@{}}Task description\\ Historical interactions\\ Context information\end{tabular}} & \multirow{3}{*}{PAP-REC~\cite{li2024pap}} & \multirow{3}{*}{T5} & \multirow{3}{*}{Free} & \multirow{3}{*}{Beauty, Sports, Toys} & \multirow{3}{*}{E-commerce} \\
 &  &  &  &  &  &  \\
 &  &  &  &  &  &  \\ \cline{2-7} 
 & \multirow{4}{*}{\begin{tabular}[c]{@{}l@{}}Task description\\ Historical interactions\end{tabular}} & \multirow{2}{*}{VIP5~\cite{geng2023vip5}} & \multirow{2}{*}{T5-small} & \multirow{2}{*}{Free} & \multirow{2}{*}{Clothing, Sports, Beauty, Toys} & \multirow{2}{*}{E-commerce} \\
 &  &  &  &  &  &  \\ \cline{3-7} 
 &  & \multirow{2}{*}{UniMAP~\cite{wei2023towards}} & \multirow{2}{*}{Redpajama-3B} & \multirow{2}{*}{Free} & \multirow{2}{*}{\begin{tabular}[c]{@{}c@{}}Baby, Beauty, Clothing\\ Grocery, Sports, Toys, Office\end{tabular}} & \multirow{2}{*}{E-commerce} \\
 &  &  &  &  &  &  \\ \hline
\multirow{4}{*}{Codebook} & \multirow{2}{*}{\begin{tabular}[c]{@{}l@{}}User profile\\ Historical interactions\end{tabular}} & \multirow{2}{*}{TIGER~\cite{rajput2023recommender}} & \multirow{2}{*}{Transformer-based model} & \multirow{2}{*}{Free} & \multirow{2}{*}{Sports, Beauty, Toys} & \multirow{2}{*}{E-commerce} \\
 &  &  &  &  &  &  \\ \cline{2-7} 
 & \multirow{2}{*}{\begin{tabular}[c]{@{}l@{}}Task description\\ Historical interactions\end{tabular}} & \multirow{2}{*}{LC-Rec~\cite{zheng2023adapting}} & \multirow{2}{*}{LLaMA-7B} & \multirow{2}{*}{Trie} & \multirow{2}{*}{Instruments, Arts, Games} & \multirow{2}{*}{E-commerce} \\
 &  &  &  &  &  &  \\ \hline
\multirow{2}{*}{Multi-facet} & \multirow{2}{*}{\begin{tabular}[c]{@{}l@{}}Task description\\ Historical interactions\end{tabular}} & \multirow{2}{*}{TransRec~\cite{lin2023multi}} & \multirow{2}{*}{BART-large, LLaMA-7B} & \multirow{2}{*}{FM-index} & \multirow{2}{*}{\begin{tabular}[c]{@{}c@{}}Beauty, Toys\\ Yelp\end{tabular}} & \multirow{2}{*}{\begin{tabular}[c]{@{}c@{}}E-commerce\\ Restaurant\end{tabular}} \\
 &  &  &  &  &  &  \\ \hline
\multirow{21}{*}{\begin{tabular}[c]{@{}c@{}}Item's textual \\ metadata\end{tabular}} & \multirow{3}{*}{\begin{tabular}[c]{@{}l@{}}User profile\\ Historical interactions\\ Context information\end{tabular}} & \multirow{3}{*}{M6-Rec~\cite{cui2022m6}} & \multirow{3}{*}{M6} & \multirow{3}{*}{Free} & \multirow{3}{*}{TaoProduct} & \multirow{3}{*}{E-commerce} \\
 &  &  &  &  &  &  \\
 &  &  &  &  &  &  \\ \cline{2-7} 
 & \multirow{5}{*}{\begin{tabular}[c]{@{}l@{}}Task description\\ Historical interactions\end{tabular}} & \multirow{2}{*}{BIGRec~\cite{bao2023bi}} & \multirow{2}{*}{LLaMa-7B} & \multirow{2}{*}{Free} & Games & E-commerce \\
 &  &  &  &  & MovieLens25M & Movie \\ \cline{3-7} 
 &  & \multirow{3}{*}{LMRecSys~\cite{zhang2021language}} & BERT-Base, GPT2-Small & \multirow{3}{*}{Free} & \multirow{3}{*}{MovieLens1M} & \multirow{3}{*}{Movie} \\
 &  &  & GPT2-Medium, GPT2-Large &  &  &  \\
 &  &  & GPT2-XL &  &  &  \\ \cline{2-7} 
 & \multirow{6}{*}{\begin{tabular}[c]{@{}l@{}}Task description\\ Historical interactions\\ External knowledge\end{tabular}} & \multirow{3}{*}{NIR~\cite{wang2023zero}} & \multirow{3}{*}{GPT-3} & \multirow{3}{*}{Free} & \multirow{3}{*}{MovieLens100K} & \multirow{3}{*}{Movie} \\
 &  &  &  &  &  &  \\
 &  &  &  &  &  &  \\ \cline{3-7} 
 &  & \multirow{3}{*}{RecRanker~\cite{luo2023recranker}} & \multirow{3}{*}{LLaMA2-7B} & \multirow{3}{*}{Free} & \multirow{3}{*}{\begin{tabular}[c]{@{}c@{}}MovieLens100K, MovieLens100M\\ BookCrossing\end{tabular}} & \multirow{3}{*}{\begin{tabular}[c]{@{}c@{}}Movie\\ Book\end{tabular}} \\
 &  &  &  &  &  &  \\
 &  &  &  &  &  &  \\ \cline{2-7} 
 & \multirow{3}{*}{\begin{tabular}[c]{@{}l@{}}Task description\\ Historical interactions\\ User profile\end{tabular}} & \multirow{3}{*}{InstructRec~\cite{zhang2023recommendation}} & \multirow{3}{*}{Flan-T5-XL} & \multirow{3}{*}{Free} & \multirow{3}{*}{Games, CDs} & \multirow{3}{*}{E-commerce} \\
 &  &  &  &  &  &  \\
 &  &  &  &  &  &  \\ \cline{2-7} 
 & \multirow{4}{*}{\begin{tabular}[c]{@{}l@{}}Task description\\ Historical interactions\end{tabular}} & \multirow{2}{*}{Rec-GPT4V~\cite{liu2024rec}} & GPT4-V & \multirow{2}{*}{Free} & \multirow{2}{*}{Sports, Clothing, Beauty, Toys} & \multirow{2}{*}{E-commerce} \\
 &  &  & LLaVA-7B, LLaVA-13B &  &  &  \\ \cline{3-7} 
 &  & \multirow{2}{*}{DEALRec~\cite{lin2024data}} & \multirow{2}{*}{LLaMA-7B} & \multirow{2}{*}{Free} & Games, Book & E-commerce \\
 &  &  &  &  & MicroLens-50K & Micro-video \\ \hline
\end{tabular}
}}
\label{tab:summary_recommendation}
\end{table*}

\textbf{Multi-facet identifier}~\cite{lin2023multi}.
To overcome the issues in previous identifier strategies, the multi-facet identifier is proposed. 
Multi-facet identifier aims to pursue both semantics and distinctiveness while mitigating the misalignment between semantics correlation and interaction correlation. 
While incorporating semantics (\textit{e.g.,} item title) exploits the world knowledge encoded in generative models, utilizing unique numeric IDs ensures distinctiveness for capturing the essential collaborative information. 
Additionally, to avoid the lengthy issue of textual metadata, TransRec~\cite{lin2023multi} allows the generation of the substring of metadata. 
The utilization of substring follows the one-to-many corresponding as discussed in Section~\ref{sec:document_identifiers}, thus might decrease the inference efficiency.

\textbf{Item identifier summary}. 
We summarize the characteristics of different types of item identifiers in Table~\ref{tab:recommendation_identifier} from different aspects, including semantics, distinctiveness, update, and the involved training process. 
From the summary, we can find that 
1) Incorporating semantics enables better utilization of world knowledge in generative language models and easier identifier update. 
This can contribute to improved generalization and practicality in real-world deployments.
2) Codebook and multi-facet identifiers achieve both semantics and distinctiveness, showing the potential to leverage semantics in pre-trained generative language models and learn collaborative information from user-item interactions. 
Nevertheless, while codebook requires additional item-to-identifier training and auxiliary alignment to endow the generated identifier with the semantics in language models, multi-facet identifier is naturally advantageous to leverage both numeric ID and descriptions for improved generative recommendation. 

\subsection{Training} 
Training the generative recommender models on the recommendation data involves two main steps, \ie textual data construction and model optimization. The textual data construction converts the recommendation data into samples with textual input and output, where the choice of input and output depends on the learning objectives. 
While most of the methods can directly construct the textual data based on the pre-defined item identifier, codebook-based methods necessitate the \textbf{item-to-identifier} training prior to textual data construction.  
The item-to-identifier training typically utilizes RQ-VAE to map the item content representation into the quantized code words~\cite{rajput2023recommender} as the item identifier. 

As for model optimization, literature typically utilizes generative training in language modeling~\cite{lewis2019bart} to optimize the model. 
Specifically, given the constructed samples with the textual input and output, the generative training maximizes the log-likelihood of the target output tokens conditioned on the input. 
According to the learning objectives, we divide the generative training into two groups for generative recommendation. 
To recommend the item relevant to the user preference, 1) \textbf{user-to-identifier} training is employed for generative models to learn the matching ability. 
For each constructed training sample, the input is the user formulation, and the output is the next-item identifier. 
The user-to-identifier training plays a crucial role in the generative recommendation and is utilized across all generative recommendation methods for item retrieval. 

As for the methods that utilize codebooks to learn semantic-aware item identifiers, 
they might suffer from the semantic gap between the quantized codewords and the semantics in natural language. 
To combat this issue, 
2) \textbf{auxiliary alignment} is additionally used by the methods to strengthen the alignment between the item content and the item identifier. 
To achieve the alignment, extra training sample construction is required, which can be broadly divided into two groups. 
(i) \textit{Content-to-identifier} or \textit{identifier-to-content}. For each constructed training sample, the input-output pair comprises the identifier and textual content of the same item, each serving interchangeably as input or output. 
In addition to the item-wise alignment, (ii) \textit{user-to-content} is another strategy to implicitly align the item identifier and the item content by pairing the user formulation with the next-item content~\cite{zheng2023adapting}. 
Despite the effectiveness of various training strategies in adapting the generative models to recommendation tasks, the training costs are usually unaffordable, especially for LLMs such as LLaMA. 
To improve training efficiency, recent efforts have focused on model architecture modification~\cite{mei2023lightlm} and data pruning for LLM-based recommendation~\cite{lin2024data,luo2023recranker}. 

\begin{figure*}[t]
\setlength{\abovecaptionskip}{0.2cm}
\setlength{\belowcaptionskip}{-0cm}
\centering
\includegraphics[scale=0.52]{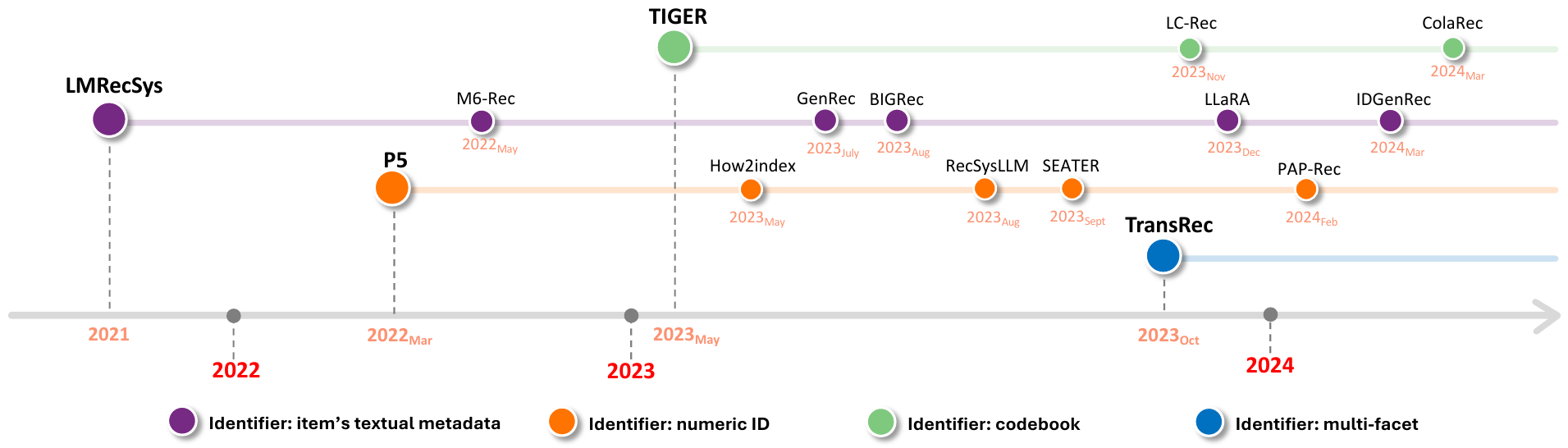}
\caption{{Major milestones in the development of generative recommendation regarding item identifier.}}
\label{fig:timeline_recommendation}
\end{figure*}

\subsection{Inference}
To achieve item recommendation, the generative models perform generation grounding during the inference stage~\cite{lin2023multi}. 
Given the user formulation in natural language, the generative models first generate the item identifier autoregressively via beam search. 
Here, we divide the generation into two types, \ie \textbf{free generation} and \textbf{constrained generation}. 
As for the free generation, for each generation step, the model searches over the whole vocabulary and selects the top-$K$ tokens with high probability as the subsequent input for the next step's generation. 
However, searching over the whole vocabulary would probably result in out-of-corpus identifier~\cite{li2023large,hua2023index}, making the recommendation invalid. 

To address this problem, early studies utilize exact matching for grounding, which conducts free generation and simply discards the invalid identifier. 
Nevertheless, they still have poor accuracy caused by the invalid identifier, especially for the textual metadata-based identifier. 
To improve the accuracy, BIGRec~\cite{bao2023bi} proposes to ground the generated identifier to the valid items via L2 distance between the generated token sequences' representations and the item representation. 
As such, each generated identifier is ensured to be grounded to valid item identifiers.

In the same period, constrained generation is also explored in generation grounding~\cite{chu2023leveraging,hua2023index,lin2023multi,rajput2023recommender,zheng2023adapting}.
~\cite{chu2023leveraging} and~\cite{hua2023index} propose to utilize Trie for constrained generation, where the generated identifier is guaranteed to be valid identifiers. 
However, Trie strictly generates the valid identifier from the first token, where the accuracy of the recommendation depends highly on the accuracy of the first several generated tokens. 
To combat this issue, TransRec~\cite{lin2023multi} utilizes FM-index to achieve position-free constrained generation, which allows the generated token from any position of the valid identifier. 
The generated valid tokens will then be grounded to the valid identifiers through aggregations from different views.

In addition to the typical recommendation that requires valid generation to recommend the existing items to the users, 
another research direction capitalizes on the generative capabilities of models to create entirely new item~\cite{xu2024difashion,wei2023towards,cui2022m6}. 
For example,~\cite{xu2024difashion} generates personalized outfits, which can serve as guidance for fashion factories. 
Consequently, within this research line, free generation is employed, enabling recommender systems to fully exploit the generative potential.

\subsection{Summary}

\textbf{Methods summary}. 
We summarize current generative recommendation methods in Table~\ref{tab:summary_recommendation}, from the aspects of user formulation, item identifier, backbone, generation, dataset, and recommendation domain. 
1) User formulation. Different components in user formulation have been discussed in Section~\ref{sec:user_formulation}. It is noticed that existing methods usually incorporate task description and user's historical interactions, while some methods also leverage user profile, context information, and external knowledge as additional components to formulate user. 
2) Item identifier. We summarize the characteristics of item identifiers in Table~\ref{tab:recommendation_identifier}, where each type of identifier meets different criteria of the item identifier as discussed in Section~\ref{sec:rec_identifier}. 
In addition, as shown in Table~\ref{tab:summary_recommendation}, we can find that methods that employ LLMs of larger size would usually utilize textual metadata as item identifier. 
This is to leverage the rich world knowledge encoded in the LLMs, for a better understanding of user behavior and item characteristics. 
With the exploration on various recommendation datasets across different domains, generative recommendation methods show great applicability and generalization ability for real-world applications. 

\textbf{Timeline summary}. 
As discussed in Section~\ref{sec:rec_overview}, item identifier is a pivotal component in generative recommendation. 
To elucidate the evolution of item identifiers in the context of generative recommendation, we provide a brief timeline to outline significant milestones across four distinct types of identifiers. 
For each type of identifier, we highlight the first work and enumerate several subsequent endeavors for further improvements in another dimension such as training strategies. 
The earliest effort in generative recommendation is LMRecSys, which employs the generative paradigm for recommendation and utilizes the item's textual metadata, \ie title, as the item identifier. 
In the year 2022, P5 introduces the numeric ID identifier and proposes a unified generative recommendation framework with multi-task training. 
Subsequently, the year 2023 has witnessed various investigations of improved numeric ID-based identifiers, such as RecSysLLM with masked language modeling~\cite{chu2023leveraging}, and SEATER with tree-based numeric IDs~\cite{si2023generative}. 
In late 2023,~\cite{lin2023multi} introduces multi-facet identifier to pursue both semantics and distinctiveness. 
It is highlighted that from year 2023 onwards, especially after the birth of ChatGPT, generative recommendation has garnered increased attention. 
In this period, there has been extensive research work to enhance generative recommendation with the four different identifier types, including training strategies (\eg GenRec~\cite{ji2023genrec} and BIGRec~\cite{bao2023bi}), user formulation (\eg InstructRec~\cite{zhang2023recommendation}), constrained generation (\eg TransRec~\cite{lin2023multi}), and training efficiency (\eg DEALRec~\cite{lin2024data}).

\subsection{LLMs for Recommendation beyond Generative Recommendation}
\textbf{LLMs as feature extractors}. 
In addition to the generative recommendation that autoregressively generates the item identifier, 
another concurrent line of research work focuses on the utilization of LLMs in data representation~\cite{wang2024rethinking,ren2023representation,lyu2023llm}. 
Two approaches are commonly used in this line of work. 1) Utilization of LLMs to obtain augmented features for traditional recommender models~\cite{xi2023towards,luo2024kellmrec,boz2024improving,sun2024large}, and
2) incorporation of a linear projector at the last token to predict the probability scores of all items~\cite{li2023e4srec,guo2024integrating,wang2024rethinking}, which is equivalent to using the LLMs' hidden states as user representation. 
Although item identifier is not explicitly utilized, it is highlighted that this line of work constructs user formulation to obtain the user representation or augmented features. 
We also include the related work into the discussion of user formulation in Section~\ref{sec:user_formulation}. 

\vspace{3pt}
\textbf{LLMs for CTR tasks}.
While extensive effort has been made to explore next-item generation, researchers have also investigated utilizing LLMs for the Click Through Rate (CTR) task.  
The CTR task aims to predict the user-item interaction in a pointwise manner~\cite{dai2023uncovering}, \ie the input is the information of the user and the target item. 
To leverage LLMs for CTR tasks, existing work usually takes the user formulation and the information of target item as the input, and the target output will be set as ``yes'' or ``no'', for the positive and negative samples, respectively. 
During inference, these methods will perform softmax on the two tokens at the output layer, ``yes'' and ``no'', and take the probability score of ``yes'' as the final prediction score~\cite{bao2023tallrec,zhang2023collm}. 
In this survey, we mainly focus on the next item generation because of its practical promise in industrial recommender systems. 
It has the potential to reduce the typical multiple-stage item ranking into one stage, \ie directly generating items to recommend. 
As for the LLMs for CTR tasks, we also discuss them in user formulation in Section~\ref{sec:user_formulation}.

\section{Discussion}
\subsection{Difference Between Generative Search and Recommendation}
In the preceding sections, we primarily outline the commonalities between generative search and generative recommendation, and summarize a universal framework to present the current works. However, given the distinct nature of their tasks, there are also numerous points of differentiation between the two and unique challenges to address, respectively.

\begin{figure*}[t]
\setlength{\abovecaptionskip}{0.1cm}
\setlength{\belowcaptionskip}{-0cm}
\centering
\includegraphics[scale=0.82]{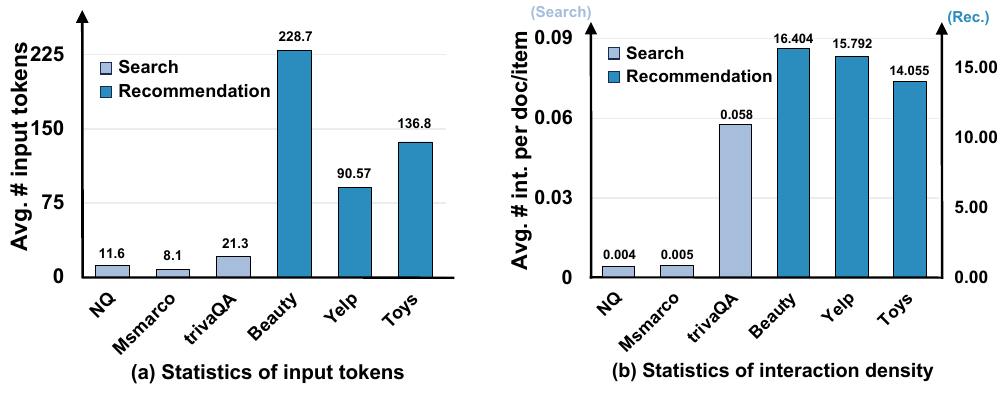}
\caption{{Statistics of six representative datasets in search and recommendation. ``Int.'' denotes ``Interactions'' and ``Rec.'' denotes ``Recommendation''. The results are based on the LLaMA2 tokenizer~\cite{touvron2023llama}. The user formulation for generative recommendation includes task descriptions and the user's historical interactions with item titles.}}
\label{fig:dataset}
\end{figure*}

\textbf{Varied input length of generative search and recommendation}. 
Generative search typically involves short queries and minimal additional processes, while generative recommendation requires a crucial "user formulation" step. This presents unique challenges for generative recommendation. Firstly, it is necessary to convert user information, such as task descriptions, historical interactions, and user profiles, into a textual sequence that preserves the original information as much as possible. Secondly, the user formulation often involves lengthy inputs. We statistics the average input length for generative search and recommendation on typical datasets, respectively. As shown in Figure~\ref{fig:dataset} (a), the input tokens for generative recommendation are significantly greater than those for generative search. The lengthy input in generative recommendation demands significant computing resources for training and poses inference difficulties, particularly with large language models~\cite{lin2024data}. Consequently, the generative recommendation must confront the challenges posed by the extensive length of the input, which is different from generative search.

\textbf{Differing interaction density in generative search and recommendation}. One important distinction between search and recommendation tasks is the difference in interaction density. In search datasets, only a portion of documents are labeled with queries, while almost all items are interacted with by users (except for cold-start items). According to the statistics in Figure~\ref{fig:dataset} (b), each document is associated with less than one query, while items typically have more than ten interactions on average. As previously mentioned, the generative paradigm involves memorizing documents or items in their parameters and then generating them based on input queries or users. In recommendation, the high interaction density ensures that each item can be trained, while most documents are not exposed to the generative search model. Consequently, generative recommendation can easily achieve comparable performance with traditional recommendation methods, whereas generative search struggles with low interaction density. The low interaction density presents unique challenges for generative search.

\textbf{Different ``semantic'' meanings in generative search and recommendation}. 
In search, the relevance between a query and a document is closely tied to their semantic similarity. However, for recommendation, an item's content is less significant compared to search, while collaborative information holds greater importance. In essence, there are distinct ``semantic'' implications in search and recommendation, leading to different requirements for identifiers in generative search and recommendation. In generative search, identifiers must accurately represent the document's content, while in generative recommendation, they should emphasize the item's collaborative information. For instance, ``learning to tokenize'' a document's content is effective in generative search~\cite{sun2023learning}, but additional collaborative information must be incorporated during the tokenization stage in generative recommendation. This may be an expected research direction to further explore in generative recommendation.

\subsection{Open Problems in Generative Search and Recommendation}
\noindent$\bullet\quad$\textbf{Document and item update in LLMs}.
After being trained on query-doc and use-item pairs, the LLM is capable of retrieving documents or recommending items to users by generating the corresponding identifiers. However, this functionality is dependent on the LLM having been trained to memorize the associations between the documents or items and their identifiers. Consequently, the LLM is hard to retrieve or recommend new documents and items that it has not encountered during training. While retraining the LLM on newly added documents or items is a potential solution, it necessitates a significant amount of computing resources. Given the vast number of new documents and items that are introduced daily in search and recommendation systems, it is crucial to develop effective and efficient methods for updating the LLM's memory in order to ensure optimal generative search and recommendation performance~\cite{chen2023continual,mehta2022dsi++}.

\noindent$\bullet\quad$\textbf{Multimodal and cross-modal generative search and recommendation}.
Only textual information cannot satisfy users' various information needs well. To enhance the search experience, it is crucial to incorporate multimodal resources such as images, tables, audio, and videos alongside textual documents. Similarly, recommendations should encompass various types of content, like micro-videos, which necessitate a comprehensive understanding of multimodal elements. Unfortunately, the current generative paradigm predominantly relies on language models and falls short in addressing multimodal information-seeking scenarios. These scenarios include retrieving images, videos, or audio based on a query, performing generative multimodal retrieval by retrieving web pages containing both text and images, and providing multimodal recommendations that incorporate diverse features. Some recent works~\cite{li2024generative, geng2023vip5, long2024generative} have made an initial attempt, and further works are anticipated.

\noindent$\bullet\quad$\textbf{In-context learning for generative search and recommendation}.
One notable advantage of Language Models (LLMs) is their ability to learn in-context. With just a few examples (few-shot) or even no examples (zero-shot) provided in the prompt, LLMs can effectively solve specific tasks without the need for fine-tuning. In-context learning plays a crucial role in enabling LLMs to encompass a wider range of tasks within the same generative paradigm. However, current generative search and recommendation methods still rely on fine-tuning LLMs using domain-specific data, including query-document pairs and user-item pairs. As the size of LLMs increases, the process of fine-tuning becomes computationally expensive. Consequently, the challenge lies in finding ways to incorporate LLMs into search and recommendation domains using zero-shot or few-shot approaches, thereby reducing the need for extensive fine-tuning.

\noindent$\bullet\quad$\textbf{Large-scale recall}.
Regardless of whether the objective is to provide users with a list of documents or items, generative search and recommendation methods typically employ beam search in autoregressive decoding to generate such lists instead of just a single result. To illustrate, if the beam search has a size of $k$, generative search and recommendation systems can produce a list containing $k$ elements. However, the efficiency of autoregressive generation diminishes as the beam size $k$ increases. Consequently, current generative search and recommendation systems are unable to achieve large-scale recall due to this limitation. Therefore, it is necessary to explore additional decoding strategies to enhance the performance of generative search and recommendation methods.

\begin{figure}[t]
\setlength{\abovecaptionskip}{0.1cm}
\setlength{\belowcaptionskip}{-0.5cm}
\centering
\includegraphics[scale=1.2]{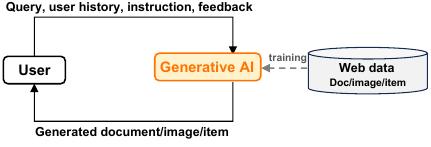}
\caption{{Illustration of next information-seeking paradigm, \ie content generation.}}
\label{fig:content_generation}
\end{figure}

\subsection{Envision Next Information-seeking Paradigm: Content Generation}
Search and recommendation systems aim to fulfill users' information needs by retrieving or recommending a list of documents or items from a finite set. In previous discussions, we have explored paradigm shifts in search and recommendation, particularly focusing on the latest generative paradigm. Regardless of whether the approach is based on deep learning or generative learning, the ultimate goal remains the same: matching the available content (documents or items) to users. However, there has been a recent surge in the development of generative AI and models, which has introduced a novel way for users to obtain information from the Web. Instead of merely matching existing content, generative AI models now have the capability to directly generate content. This means that the generated content, such as a document, may not have previously existed on the Web. 

\noindent$\bullet\quad$\textbf{Search engines vs generative language models}. Search engines, such as Google\footnote{\url{https://www.google.com/}.}, are widely used text retrieval applications that can efficiently retrieve relevant web pages from billions of websites based on user queries. Generative language models like ChatGPT~\footnote{\url{https://chat.openai.com/}.} and Gemini~\footnote{\url{https://gemini.google.com/}.} are trained on vast amounts of web documents to imbibe knowledge into their parameters. These models can then generate responses to user queries based on the instructions provided.

As a result, users now have the option to input their queries directly into generative language models to seek information. In comparison to search engines, generative language models offer several advantages. 1) They can provide more accurate responses, as they are not limited to presenting an entire web page. 2) They can summarize content from multiple pages, which is particularly useful for complex queries. 3) Additionally, generative language models can understand users' information needs in multi-turn conversations. However, generative language models do face certain challenges. One such challenge is the issue of information updating. Since these models are not trained on the latest documents, they may not be able to answer queries related to up-to-date information. Consequently, they may lack the corresponding knowledge. Furthermore, generative language models are prone to generating incorrect responses, a phenomenon known as hallucination~\cite{ji2023survey}. Overall, while generative language models offer unique advantages in information retrieval, they also have limitations, including information updating and hallucinations, that need to be addressed.

\noindent$\bullet\quad$\textbf{Image search vs image generation}. People have become accustomed to searching for images on the internet to fulfill their visual requirements. However, the advancements in image generation, such as the GAN~\cite{goodfellow2020generative} and diffusion models~\cite{croitoru2023diffusion}, now enable individuals to directly acquire the desired images through generation rather than relying on web searches. This is especially beneficial for creative endeavors, as image generation has the potential to produce distinctive and unique visuals. Compared to image search, image generation models offer more personalized services and can create unique images based on users' specific requirements. However, these models are also prone to the issue of generating images that do not adhere to physical rules. This means that the generated images may not accurately represent real-world objects or scenes.

\noindent$\bullet\quad$\textbf{Item generation in recommendation}. 
Traditional recommendation systems retrieve existing items from item corpus, which may not always meet users' diverse information needs. 
Nevertheless, the boom of generative models offers a promising solution to supplement human-generated content in the traditional recommendation and foster personalized AI-Generated Content (AIGC)
Specifically, item generation in recommendation can be achieved in two approaches. 
1) Content repurposing, which edits the existing items tailored to individual user preference for personalized recommendation~\cite{wang2023generative}, \eg transfer a micro-video in a cartoon style. 
2) Content creation, which generates personalized new items to satisfy user-specific preference and intent~\cite{cui2022m6,xu2024difashion,wei2023towards,shen2024pmg}. 
By leveraging these two approaches, personalized item generation enhances information seeking by providing personalized open-world content that aligns closely with individual user preferences.

\section{Conclusion}
Bridging the semantic gap between two matching entities is a fundamental and challenging issue in search and recommendation. In this survey, we provide an overview of three paradigms aimed at solving this core research problem from a unified perspective. Our focus is primarily on the recent generative paradigm for search and recommendation, known as generative search and recommendation. We provide a comprehensive overview of a unified framework for generative search and recommendation, and discuss the current research contributions within this framework. In addition to categorizing the current works, we offer valuable insights into the generative paradigm, including an analysis of the strengths and weaknesses of various designs, such as different identifiers, training methods, and inference approaches. By adopting a unified view of search and recommendation, we highlight the commonalities and unique challenges of generative search and recommendation. Furthermore, we engage in a thorough discussion about the future outlook of the next information-seeking paradigm and the open problems in this field.
\clearpage

\bibliographystyle{ACM-Reference-Format}
\bibliography{sample-base}


\end{document}